# Perspectives and solutions towards intelligent ambient assisted living systems

Hong Sun[1], Vincenzo De Florio[2]

[1] Dedalus HealthCare, Gent, Belgium

[2] Evolution, Complexity and COgnition research group

Center Leo Apostel, Vrije Universiteit Brussel, Belgium

{hong.sun@dedalus-group.com, vincenzo.de.florio@vub.be}

**Abstract:**

The population of the elderly people has kept increasing rapidly over the world in the past decades. Solutions that are able to effectively support the elderly people to live independently at their home are thus urgently needed. Ambient assisted living (AAL) aims to provide products and services with ambient intelligence to build a safe environment around people in need. With the high prevalence of multiple chronic diseases, the elderly people often need different levels of care management to prolong independent living at home. An effective AAL system should provide the required clinical support as an extension to the services provided in hospitals. Following the rapid growth of available data, together with the wide application of machine learning technologies, we are now able to build intelligent ambient assisted systems to fulfil such a request. This paper discusses different levels of intelligence in AAL. We also introduce our solution for building an intelligent AAL system with the discussed technologies. Taking semantic web technology as its backbone, such an AAL system is able to aggregate information from different sources, solve the semantic gap between different data sources, and perform adaptive and personalized carepath management based on the ambient environment.

# 1. Introduction

As well known, the increase of the elderly population is a significant demographic and socioeconomic concern. Already very high in 2019 (9% of people are aged 65 or over), projections foresee a world-wide rate of about 16% in 2050 and of about 25% when restricting the focus on Europe and North America [58]. As a consequence of this threat, several actions have been planned, starting with the United Nation's first World Assembly on Ageing in 1982. A major concept was introduced in 2005: Ambient Assisted Living (AAL) [4, 5]. AAL aims at assisting the aging people living well in their own home by increasing their autonomy and self-confidence. Following this concept, the AAL programme [6] was built up. Said program has funded over 220 projects since 2008 and is still active today.

The introduction of AAL is a consequence of this state of affairs: rather than purely relying on human assistance, which is scarcer because of global aging, AAL capitalizes on the ever-increasing performance and pervasiveness of computer and networking systems in order to supply the aging population with some form of "smart" assistance so as to increase the autonomy, safety, and comfort of the aging ones. Already in our paper [7] we discussed the promises and challenges of this family of approaches, also highlighting the need for an evolution of the concept of AAL. This need was confirmed by the emergence of new and more advanced forms of intelligence in the past decade. Moreover, since the elderly populations are often suffering from multiple chronic diseases, it is important that an intelligent AAL system could provide clinical support as an extension of the care support provided in hospitals. The AAL systems are evolving to support more advanced application in meeting such a requirement. We consider that the evolution of AAL is in general following the three phases below:

In the first phase, AAL was merely intended as the establishment of a ICT-backed "safety net" around the patient. This phase is characterized by the wide application of different types of sensors, and the AAL system detects dangerous situations around the person in assistance, such as a fall event.

A second phase was characterised by the rapid adoption of Electronic Healthcare Records (EHR) in combination with sensor data from smart body devices such as smart watches and the availability of health data, in structured or unstructured format. This led to predictive AAL services, able to anticipate the need for intervention thanks to predictions backed by machine learning solutions.

The third phase, which we are currently experiencing, takes a step further and formulates AAL approaches able not just to predict potentially dangerous events but also to formulate and orchestrate solutions to deal with the foreseen events.

We observe that a major aspect shared by AAL systems in the first two phases is that they focus on detecting/predicting risk events. AAL systems in the third phase, on the other hand, are characterized by prescriptive intelligence, in that they provide guidance to resolve risk events, or provide basic carepath management. In order to build an intelligent AAL system that is able to utilize the information, resources, and the predictive and prescriptive intelligence from different applications, it is important to first build a foundation layer that enables the communication and understanding between different data and applications. We deem this foundation layer as a type of descriptive intelligence.

In this paper, we discuss the application of descriptive, predictive and prescriptive intelligences in the AAL systems. We also discuss the challenges and provide our considerations in view of building these three types of intelligence. Lastly, we give our visions in combining the descriptive, predictive and prescriptive intelligence to build an intelligent AAL system. We use a scenario of detecting and resolving a fall event to demonstrate the capability of such an intelligent AAL system.

## 2. Types of intelligence required by an ambient assisted living system

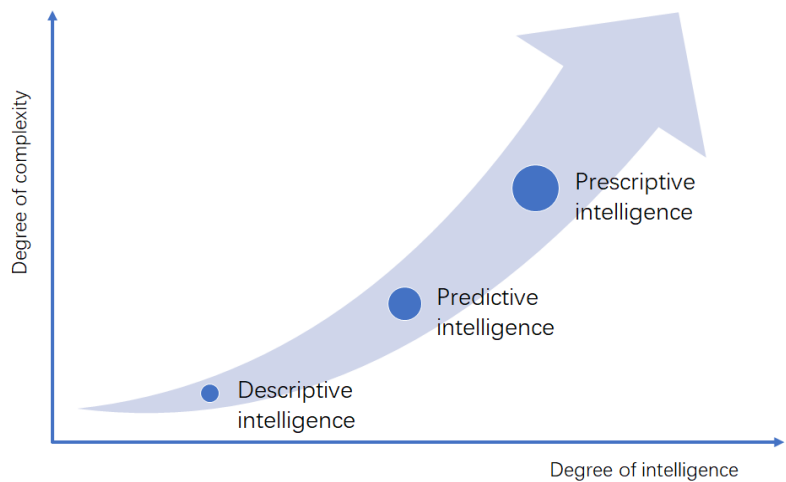

Figure 1. Level of intelligence in ambient assisted living

Ambient assisted living systems are the result of a complex interplay of players, both human and technological, that provides a wide spectrum of services. The wide application of smart devices, the easier reach to the EHR data, and the advances of machine learning technology open new opportunities to build more intelligent AAL systems. The ever more advanced features of those systems introduce ever bigger new challenges, such as the ability to cope effectively with more complex situations and to provide personalized support to elderly people affected by comorbidity. An intelligent AAL system needs to perform tasks once considered the exclusive territory of man: understanding ongoing situations regarding human beings in need; predicting trends; assessing potential risks; coordinating the action of multiple players in such a way as to avoid or minimize risks. Different types of intelligence are therefore required to build an intelligent assisted living system.

Several authors provided characterizations of intelligence in AAL or other domains from a perspective of data analytics. Gingras [3] proposed to build a scalable and flexible AAL system with data analytics workflow divided into four stages: sensor layer, data preprocessing, data processing pipelines, and knowledge and insight. Deka [1] and Gensinger [2] described data analytics as descriptive analytics, predictive analytics and prescriptive analytics in the domain of big data analytics and healthcare analytics. Our definition of the levels of intelligence in AAL also classifies them into a descriptive, a predictive, and a prescriptive component:

A first requirement is descriptive intelligence. This is the ability to acquire awareness of the current status, which in turn requires the ability to aggregate and "understand" data produced by different actors. A second required feature is predictive intelligence: the ability to make use of the accrued context information to predict risk events. Lastly, it requires prescriptive intelligence -- based on context awareness and predicted risks, the system adapts and coordinates relevant players to guarantee the well-being of the person to assist. In the rest of this section we discuss these three types of intelligence; we detail their focuses, also highlighting specific differences with respect to descriptive, predictive and prescriptive analytics; and we describe our solutions for bringing each of these intelligences in AAL systems.

## 2.1 Descriptive intelligence - the ability to understand

The first level of required intelligence in AAL is descriptive intelligence. Both descriptive intelligence and descriptive analytics focus on preparing and analyzing historical data [1]. The difference is that data analytics emphasizes the visualization, reporting and identification of patterns, while descriptive intelligence focuses on linking data from different sources, and building common semantics so that data from different sources can be coherently reasoned upon.

As already mentioned, an AAL system is a complex system that involves many different players as well as data from different resources. Being able to aggregate data from different sources, and share a common semantic is crucial to provide an accurate perception of the condition of the person in assistance as well as of the context environment that person is located in. The descriptive intelligence constitutes the foundation of an intelligent AAL system: prediction and guidance capability is built upon the information provided by this layer.

Wearable devices equipped with sensors, such as smart watches, are now widely used to measure the vital signs of the wearer [10]. Most of those devices can monitor pulse rate, and recent products now support monitoring the level of oxygen saturation (SpO2) and electrocardiogram (ECG) functionality. As remarked, e.g., in [10,11,12], these devices provide data of significant value in monitoring health status. Besides monitoring vital signs, wearable devices feature different sensor technologies to assist in chronic disease management. Examples include wearable sensors for diabetes management [13] and contactless sensors for Parkinson's disease management [16]. Haque et al. [14] studied the application of using ambient devices in the healthcare domain and assisted living, and confirmed the richness of information gained through ambient sensing from hospitals and daily living spaces. Besides the clinically relevant application, other sensor technologies such as humidity sensors and co2 sensors are also playing an important role in building a smart environment to assist people's daily lives.

While the wearable and contactless sensors are providing valuable data regarding the person in assistance, as well as providing different valuable functions, it is crucial that an AAL system understand the data coming from different sensors, as well as understand the services provided by different smart devices. We consider such an ability as a fundamental aspect of an AAL system's descriptive intelligence layer.

To build predictive intelligence in AAL, it is important to connect data from different sources in an AAL system and provide explicit meanings to the data. Berners-Lee et al. [17] proposed semantic Web, a web of data that can be processed directly or indirectly by machines, as a solution to build common understandings among different data sources. The RDF Vocabulary Definition Language (RDFS) and the Web Ontology Language (OWL) provide a basis for creating ontologies that can be used to describe entities as well as the relations between entities. By publishing and connecting different data sources as linked data, it lowers the barrier to reuse, integration and application of data from multiple, distributed and heterogeneous sources [18]. Tao [7] and Ni [8] developed ontologies to model human activities and smart devices in the smart home environment. Pardo [9] developed a framework for anomaly diagnosis in smart homes based on ontology and a reasoning engine. Malek et al. [19] used IoT and big data technologies for real-time monitoring and data processing in smart home environments.

Besides the vastly increasing data from smart devices in the home environment, the usage of the EHR is also growing considerably throughout the last decade. EHRs are normally stored in the hospitals and can be accessed through EHR systems. Recent efforts have been made at government level to provide patients with access to their EHRs [21]. Compared with data from wearable devices, EHR from hospitals provides more structured data with more clinical values. More and more hospitals are now using EHRs and opening their access to patients. Aggregating patient generated health data from assistive devices with EHRs has the potential to close healthcare gaps and support personalized care. Tiase et al. [22] reviewed the progress of such integration and pointed out that such an integration is still at an early stage and efforts are required. We consider that a barrier to such an integration is the fact that the data is recorded in different formats from disparate sources, which translates into "semantic gaps" between these data.

Using common standards can facilitate the data integration of the patient generated data and EHRs. El-Sappagh et al. [20] built a mobile health monitoring-and-treatment system based on the integration of the SSN sensor ontology and the HL7 FHIR standard [27] to monitor and manage type 1 diabetes mellitus. We also developed semantic data virtualization techniques to allow data from different sources into representations with domain ontology, and support different clinical research applications with data represented in a domain ontology [23].

We consider building a semantic integration framework to integrate senor data, patient generated data and EHRs as a solution to reach descriptive intelligence in AAL. Such a framework should aggregate data from different sources and achieve semantic interoperability by relying on a set of common standards and semantics. This would allow integrating data from different sources and having their meanings understood by different machinery applications in support of more advanced predictive and prescriptive intelligence services.

In choosing common standards for semantic integration framework, we express two major requirements:

1. First, the adopted standards should be widely used and accepted and cover a reasonably large domain. Many projects have put significant effort into creating ontologies to support data exchange. However, rather than contributing to a common ontology and thus facilitate knowledge sharing, those projects typically focused on limited aspects and produced domain-specific ontologies, which restricted their reuse outside of the project scope. The consequence is that, instead of facilitating knowledge sharing, they actually turned to be barriers of knowledge sharing.

2. Secondly, the standards should support data representation with a semantic web language. We propose to use semantic web technology to achieve semantic interoperability between data from different sources, and assign explicit semantics to those data to achieve descriptive intelligence. It is therefore important that the chosen common standards can provide such support.

The HL7 FHIR standard is the most favourable choice to model clinical data [27]. FHIR provides an agile and RESTful approach to healthcare information exchange [24]. It also supports representing FHIR resources as RDF graphs[1]. FHIR is proposed as the standard for health data interoperability in the US [21], and it is widely used in European countries such as Germany and Austria [25]. An ongoing effort is aiming at building an interoperable app platform that promotes building standard applications on FHIR [28].

While FHIR serves as the standard to model the clinical domain, the non-clinical domain also requires to be modelled in a formal way. Schema.org is recommended to model such non-clinical domains, such as daily activities or data from other types of sensors [26]. Schema.org is founded by Google, Microsoft, Yahoo and Yandex, and aims to create a single integrated schema to cover a wide range of topics, including people, places, events, etc. The Schema.org vocabularies are developed by an open community process, and can be used with many different encodings, including RDFa. They are now being widely used on the web and other applications to power rich, extensible experiences.

---

[1] https://www.hl7.org/fhir/rdf.html

It is also worth to mention that Schema.org also have vocabularies in the clinical domain[2]. Those vocabularies can be used to build applications in an AAL system, mostly as patient-generated clinical data. However, we also consider that in order to support predictive and prescriptive intelligence based applications, a mapping between the clinically relevant data represented with Schema.org and EHRs represented with FHIR should be established -- mostly a mapping from Schema.org to FHIR. We propose to use our semantic data virtualization technique [23] to establish such a virtual mapping so that source data can be kept in their original format in their own data repository, and only execute the mapping when needed.

## 2.2 Predictive intelligence - the ability to infer from facts

While descriptive intelligence aggregates data from different sources and assigns explicit semantics to the data, it also builds the foundation for predictive intelligence. Predictive intelligence is similar to predictive analytics since both focus on making predictions based on the aggregated data. In predictive analytics, Deka [1] considers predictive analytics is about predicting the future, and Gensinger [2] also classifies actions such as making alerts after analyzing the data as descriptive analytics rather than predictive. We define predictive intelligence with a broader scope: we consider predictive intelligence as the ability to derive further information from the existing facts, either with full confidence (e.g. through deductive reasoning) [29] or with certain probabilistic belief (e.g. through inductive reasoning or machine learning algorithms [29, 30]), regardless it is predicting an event to happen in the future or deducing new information based on known facts. Therefore, we consider making alerts after analyzing the data as a kind of predictive intelligence.

A typical scenario of predictive intelligence in AAL is fall detection[3]. In 2004, 28.7% of elderly adults in American reported falling at least once a year, 37.5% of those who fell reported at least one fall that required medical treatment or restricted their activity for at least 1 day [31]. Different sensor technologies are used to develop fall detection systems. Chaudhuri [32] reviewed fall detection systems and their use with elderly people. Their review found that the wearable devices achieved 96% accuracy in fall detection.

---

[2] https://schema.org/docs/meddocs.html
[3] It may seem confusing to classify fall detection as a form of predictive intelligence; still, we prefer to do so because such a class of methods typically makes use of an inference process.

Machine learning algorithms are now also widely used to predict risk events for clinical applications [33, 34]. Driven by the increases in computational power, the advances in machine learning techniques, and the availability of standard electronic health records (EHR), the machine learning algorithms are able to accurately predict clinical risks such as Parkinson's disease, delirium etc. that are common in the elderly population [35, 36]. Although most of such disease predictions are trained with EHRs and used in the hospital settings, it is also possible to aggregate the EHR data with the data generated in an AAL system to make predictions in the AAL domain, and improve the accuracy of the predictions. As an example, delirium is a clinical syndrome defined as an organically caused disturbance in attention and awareness over a short period of time. An important feature in delirium prediction is the activeness of daily life, which is an aspect often missing in EHRs though can be provided by an AAL system. In Parkinson's disease prediction, Del Din [37] used wearable devices to make gait analysis and predict occurrence of the disease.

It is important to mention that although the inductive reasoning, to infer a theory from data, is popular as most of the machine learning algorithms follow this approach, we should still not ignore the power of deductive reasoning: applying known theory on data to infer new information. De Florio [38] implemented context aware analysis in an AAL environment with predefined rules. Spoladore [39] applied Semantic Web Rule Language (SWRL) [40] in an AAL system to infer new information for clinical decision support. When the clinical knowledge is formally defined with explicit semantic, it is possible to deduce further knowledge with explicitly stated rules, such as calculating Body Mass Index (BMI) based on the weight and height of a person. For instance, Mohammadhassanzadeh [41] extended the knowledge coverage of medical knowledge bases for improved clinical decision support with semantic based reasoning.

We observe how the advances in machine learning technology and the widely increasing amount of health data of the past decade resulted in a significant improvement of the quality of predictive intelligence methods for AAL. We believe such a trend will be confirmed in the future, with a steady reduction of the factors limiting the development of prediction services in AAL. The main challenge in this domain is mainly about optimized utilization of the resources, which has three folds of meanings:

- First, the ability to best utilize the available data in developing prediction services. We deem that by assigning explicit semantics in the descriptive intelligence layer, in combination with the adoption of rule sets to deduce further knowledge, the available data can be better utilized. While data from different sites are represented in the same standard, it is possible to apply federated learning to train a global model with sensitive EHR data that are still located in their respective hospitals [42].

- Secondly, the ability to best utilize the efforts in developing machine learning algorithms for prediction services. Currently, the development of different clinical risk prediction services are carried out independently, with few efforts being made with the aim to reuse the development process and engineer a scalable approach to develop prediction services. An additional problem stems from the fact that most of the clinical prediction services are relying on peculiar characteristics of the training data. In other words, training data reflects a reference environment, and the performance of models built upon the training data is not guaranteed in a different environment. Rajkomar [43] used FHIR as a standard data representation of EHR, and developed a set of prediction services in the clinical domain in a scalable way. Relying on a standard data representation and prediction service development procedure, we also developed a scalable approach to build multiple clinical risk prediction applications at different sites based on model calibration at the deployment site [44]. Our lessons learned tell us that, by making use of standardized data provided by the descriptive intelligence layer, and by applying common feature processing methods, it is possible to reuse much of the effort in developing clinical risk prediction applications in AAL systems.
- Lastly, the ability to best utilize the predicted results of the various prediction services. Although there are many prediction applications developed in the AAL domain, most of them are isolated from each other. We consider prescriptive intelligence as a solution to utilize the predicted results to create more effective applications.

## 2.3 Prescriptive intelligence - the ability to guide

We define prescriptive intelligence as the ability to instruct an AAL system about how to act in the face of the detected/predicted conditions. This is in line with what Deka [1] defines as prescriptive analytics: the ability to evaluate and define new ways to operate, and to target objectives while balancing all constraints.

Prescriptive intelligence is often coupled with predictive intelligence; prescriptive AAL systems are capable of making corresponding adaptations following the detection of events or situations. Loreti [45] built an AAL system using a knowledge based complex event processing engine to detect situations such as the presence of the caregivers or movements of the person in assistance, and used reactive event calculus [59] to take plan and enact corresponding reactions. Botia [46] detected abnormal situations by analyzing incoming sensor data with a decision rule set, and took corresponding adaptations following the user's behaviour. In [47], we also investigated an AAL system on a societal level, such that when abnormal events such as falls are detected, the system takes the corresponding reaction by looking for help from neighbours.

Elderly people are often affected by chronic disease or comorbidities and require different levels of clinical support [48]. Integrated care pathways [49] aims to integrate multidisciplinary care plans. Such plans are largely influenced by the patient's clinical condition as well as by social circumstance. We consider that an intelligent AAL system should share the responsibility of providing clinical support, to be considered as the extension of clinical pathway management. Moreover, it should be endowed with prescriptive intelligence, able to analyze the consequences of the detected/predicted conditions, and to formulate guidelines or instructions.

Liu [50] reviewed the effect of smart homes on the elderly people with chronic diseases -- the effectiveness of tele-monitoring was confirmed. Nevertheless, the author also reported the side effect of introducing depression. An AAL system with prescriptive intelligence could reduce the demand for tele-monitoring, which not only saves medical resources but also alleviates the depression of the patient that is in assistance. Chi [51] developed such a chronic disease dietary consultation system using OWL-based ontologies and semantic rules, to guarantee sufficient nutrition intake for patients with chronic kidney disease.

Introducing high level prescriptive intelligence into an AAL system is a challenge with many preconditions. In order to make proper guidance, a system first needs to have an accurate picture of the context environment and of the condition of the patient, as well as it requires knowledge of the relevant healthcare and AAL domain and the possible reactions. We deem that the descriptive and predictive intelligence layers fulfil such requirements by aggregating the required information and representing them in standard formats that the system can understand.

A second challenge to the application of prescriptive intelligence in an AAL system is that for patients with comorbidity, there are multiple carepaths ongoing, as well as different ongoing daily life activities. Consequences of actions may interfere with one another, and so do the possible reactions or adaptations. It is therefore important for the prescriptive intelligence to be aware of these consequences, and guarantee that its management does not interfere with existing paths and does not conflict with other ongoing or planned actions. This is only possible when the consequences of each action are explicitly stated. We have investigated the interactions between different carepaths in the GPS4IC (GPS for Integrated Care) project [52], and developed a weighted state transition logic [53] to formalize the descriptions of actions with semantic language by explicitly describing the associated consequences. Running with path generation and path validation engine, the GPS4IC system is able to generate adaptive and personalized carepaths, as well as to detect possible conflicts between different carepaths, even if the action that would cause the conflict is not carried out yet.

Although we introduced descriptive, predictive, and prescriptive intelligence separately, these three components are often combined and mixed organically in AAL applications. In summary, the descriptive intelligence builds the foundation to understand the context environment and serves as the perception layer providing the other two levels with required information. The predictive intelligence makes detections or predictions based on the received information, and the prescriptive intelligence is often triggered by both the predictions generated with the predictive intelligence and the information updated by the descriptive intelligence. We deem that all three components are indispensable to build a truly intelligent AAL system. In the next section, we introduce our vision of building an intelligent AAL system through the cooperating action of these three intelligence components.

## 3. Towards an effective AAL system with integration of intelligences

Figure 2 shows a blueprint to build an effective AAL system with descriptive, predictive and prescriptive intelligence. The elderly people are often affected by chronic diseases and require medication or other care services. As mentioned previously, the prevalence of comorbidity is high in the elderly population, with 80% of them having three or more chronic conditions [48]. As a consequence, polymedication in the elderly is common. In Australia, almost 88% of those aged 65 years and over use at least one

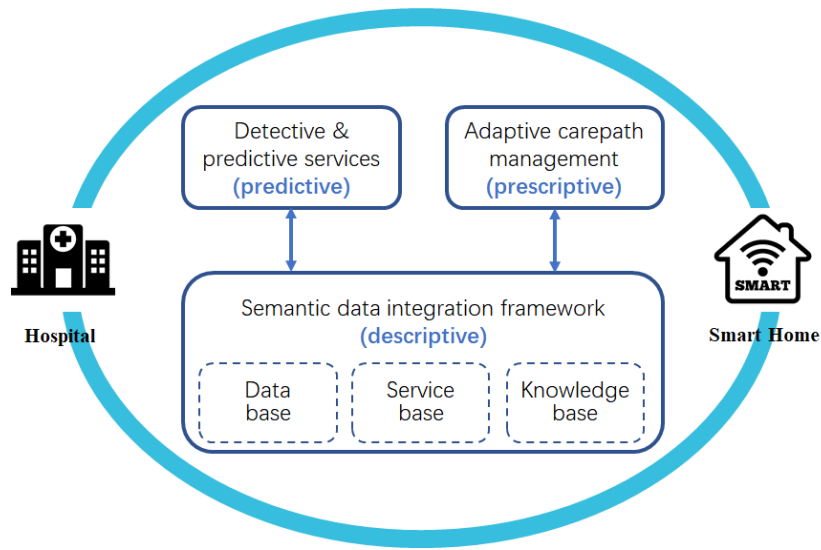

Figure 2. An effective AAL system with descriptive, predictive and prescriptive intelligence

prescription medication [54]. With increasing numbers of concurrent carepaths and medications, the risk of conflicts between different carepaths and adverse drug events increases significantly. We consider that an intelligent and effective AAL system should allow healthcare management to be extended from the hospital to the home of the patient. IoT ecosystem that connects humans and devices, and links data and applications from different resources is considered as the ideal solution to build such an AAL system. In our previous work [61, 62], we also considered the participation of humans as a key ingredient in building an effective AAL system. We consider descriptive intelligence as a means to bridge the semantic gap between the EHR data and the patient generated data in the AAL domain, in that it enriches the available data to both clinical applications in hospitals as well as assistive services at home. Properly tailored so as to detect and predict risks and undesired conditions in the home domain, predictive intelligence extends the clinical risk prediction services from the hospital to our homes. Prescriptive intelligence also extends the clinical pathway management from the hospital to the home domain, and it extends its scope from the treatment of a particular disease to the management of multiple care plans for comorbidities and daily activities. Figure 2 shows our vision to combine these three types of intelligence into an AAL system, also highlighting the interactions we envision between each other. Executional details such as calling services from a clinical pathway or executing an action following the guidance of the prescriptive intelligence are out of the scope of this figure. The three types of intelligence are built and interact as follows:

The semantic data integration framework constitutes the backbone of the intelligent AAL system. It establishes the descriptive intelligence by relying on standard data representation, and also serves as the foundation to provide inputs to the predictive and prescriptive services. At the same time, the semantic data integration framework also serves as a bridge to connect the detective and predictive services with the adaptive carepath management. The framework consists of a data base, a service base, and a knowledge base. The data base contains patient records from hospitals, as well as data generated by different sensors in the home domain. The service base contains descriptions of services provided by assistive devices, formal/informal caregivers, or even the persons in assistance themselves. Clinical activities such as taking a medication, or daily activities such as doing some physical exercise are both considered as services. The services stored in the service base are the basic elements to construct an adaptive carepath. The knowledge base stores knowledge in the clinical and AAL domain, such as calculating BMI based on weight and height, as well as the predictions generated by different prediction services, and the carepaths generated by the adaptive carepath management.

The detective and predictive services use information from semantic data integration work as inputs to detect or predict abnormal situations and events. Such detections and predictions can either be based on predefined rules, or predicted with machine learning algorithms. The detected or predicted events are saved in the semantic data integration framework, and serve as important information for adaptive carepath management of the prescriptive intelligence.

The adaptive carepath management is triggered when new information, either new abnormal events or new data, is entered. The adaptive carepath management checks whether existing carepaths/plans can still reach their respective targets given the updated information, and is also responsible for generating adaptive carepaths to solve possibly new abnormal events. A carepath consists of multiple services as described in the service base of the semantic data integration framework. A service description describes the state before and after the execution of the service, as well as the preconditions of executing the service. The process of executing a service is then considered as the process of moving from the state preceding execution to the state following execution. We use weight state transition logic [53] to carry the task of path generation and path validation. During the process of path generation and path validation, the model continuously updates its present state with the predicted future state. Such a process obeys the Markov property [55]: the conditional probability distribution of future states of the process depends only upon the present state, not on the sequence of events that preceded it. By predicting the future state, the system is also capable of detecting potential conflicts (in the future state) that are caused by actions belonging to different carepaths. Information from both data and knowledge bases are used in the process of path generation and path validation. The source code of weighted state transition logic is published on Github (https://github.com/hongsun502/wstLogic) together with a scenario of generating carepaths for colon cancer and Parkinson's disease, as well as detecting conflicts between the two paths. In the rest of this section, we exemplify the application of weighted state transition logic to generate adaptive and personalized carepaths to guide the response to a detected fall event in the AAL domain.

## 4. Scenario - planning response to resolve a detected fall event

This section uses the schedule of responses to a detected fall event as a scenario to show how the descriptive, predictive and prescriptive intelligence interact with one another and generate adaptive personalized carepaths to guide the response. The scenario is implemented with weighted state transition logic. A detailed implementation of this scenario, including e.g. the service descriptions, path generates, knowledge representations, generated paths under different context, are published in Github (https://github.com/hongsun502/AALDemo/tree/main/fall_detection)

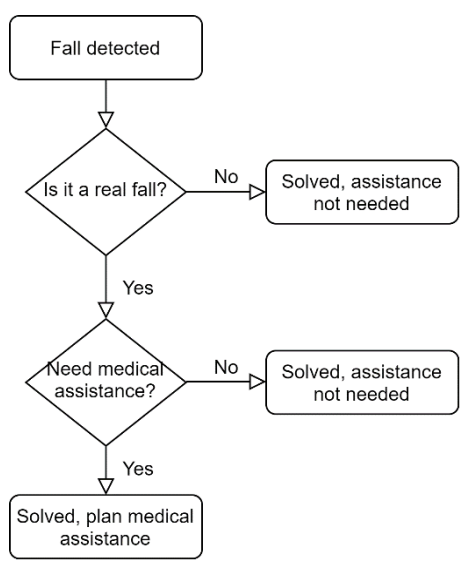

Figure 3. Simplified workflow of possible responses to a detected fall event

Figure 3 shows a simplified scenario of possible responses when a fall event is detected by a predictive intelligence component. The prescriptive intelligence plans the responses, e.g. to check whether it is an actual fall or just a false alarm -- so as to assess whether medical assistance is needed. It is not difficult to implement a decision support system following the workflow presented in Figure 3. It would be a bit more challenging to include the background information and knowledge to make inference in assisting the decision making -- a strong support of descriptive intelligence would make this task easier. A critical aspect is that the possible responses are following the fixed workflow, thus are not able to investigate different paths that may as well lead to the target. A truly intelligent AAL system should be an open system such that additional information and resources are continuously manifested themselves. Therefore, they should be integrated into the system to enlarge the 'solution horizon'. However, any static solution, such as the fixed workflow, may not utilize well new resources to explore optimized solutions.

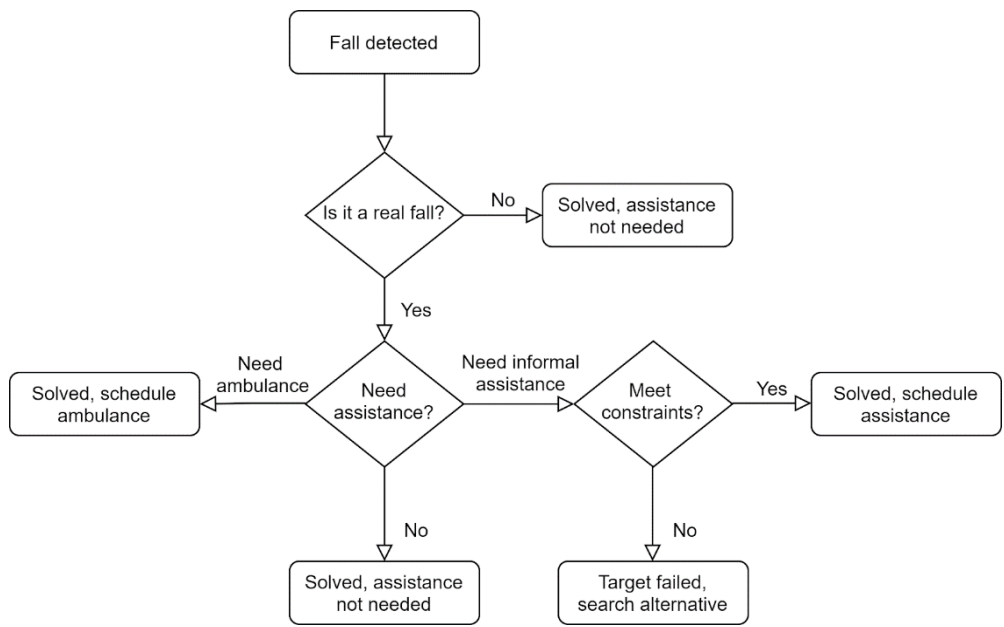

Figure 4. A workflow with extended possible responses to a detected fall event

As an example, Figure 4 shows a workflow with additional possible responses compared with the one in Figure 3. This second workflow takes more detailed analyses to schedule more detailed plans. Additional works are therefore required to adapt the fixed workflow in Figure 3 to the one presented in Figure 4. However, when new services are available or new requirements are posted the workflow presented in Figure 4 also may be recognized as a simplistic, and thus limiting, solution. In order to accommodate with the new context, the decision support system would again need to be updated to accommodate for a revision of the workflow. The above example highlights a factor limiting the effectiveness of any approach based on a fixed workflow. Given the existence of many different and continuously expanding carepaths in an AAL system, fixed workflows continuously call for new updates, and this makes it difficult to maintain them.

As a solution to the above problem, we use weighted state transition logic [53] to explore the new 'solution horizon' by automatically generating an adaptive and personalized path, which consists of a set of actions that lead from the current state to the target state. When new services are entered into the system, the system only needs to add the corresponding service descriptions, so that the services can be considered while generating a new carepath.

```
Action: Confirm fall(pos)          Action: Confirm fall(neg)          Action: Check assistance (a)    Action: Check assistance (b)
From state:                         From state:                         From state:                     From state:
   Fall is detected                    Fall is detected                    Fall is confirmed               Fall is confirmed
Target state:                       Target state:                       Target state:                   Target state:
   Fall is confirmed                   Fall is not confirmed               Ambulance is required           Non-medical assistance
   Event to be resolve                 Event is solved                                                     is required
                                                                                                       Condition:
                                                                                                          Ambulance is not required

Action: Check assistance (c)        Action: Schedule ambulance          Action: Schedule assistance
From state:                         From state:                         From state:
   Fall is confirmed                   Ambulance is required               Assistance is required
Target state:                       Target state:                       Target state:
   Assistance is not required          Ambulance arranged                  Assistance is arranged
   Event is solved                     Event is solved                     Event is solved
Condition:                                                              Condition:
   Ambulance is not required                                               Distance between provider and receiver
                                                                          Provider and receiver should be related
                                                                          Capable to provide assistance
```

Figure 5. Service descriptions (abstract descriptions)

Figure 5 shows the abstract descriptions of a set of possible actions. Each action is considered as an activity that brings changes to the state of the person in assistance. Both the state before the action, as well as the state after the action are explicitly stated. When an action is executed, the state before the action will be replaced by the state after the action -- this is the core concept of the weighted state transition logic. In situations in which the state after the action is not certain, as it is the case for the action to confirm a fall event, both possibilities are explicitly stated in two separated service descriptions. A service description may also include preconditions to the execution of the associated action. As an example, an action to schedule assistance may introduce preconditions such as "the selected helper should be capable of providing assistance", or "should be related to the person in need of assistance", or "should be within a certain distance from the person in need of assistance".

```
#Basic information location and relation          #simplified case of calculating distance
data:patient_1 care:location data:Location_A_1.   #should be caclulate by an extra service in practice
data:person_b care:location data:Location_B.      (data:Location_A_1 data:Location_A_2) care:distance 1 .
data:person_b a care:Informal_Assistant.          (data:Location_A_1 data:Location_A_3) care:distance 1 .
data:person_c care:location data:Location_A_2.    (data:Location_A_1 data:Location_B) care:distance 100 .
data:person_c a care:Informal_Assistant.
data:person_d care:location data:Location_A_3.    #Inference rule to infer relationship
data:patient_1 care:neighbour data:person_c.      {?A care:neighbour ?B.} => {?A care:relatedTo ?B.}.
data:patient_1 care:neighbour data:person_d.      {?A care:family ?B.} => {?A care:relatedTo ?B.}.
data:patient_1 care:family data:person_b.         {?A care:relatedTo ?B.} => {?B care:relatedTo ?A.}.
```

Figure 6. Background data and knowledge

Figure 6 shows some background data and knowledge to assist in the care path generation. The data is expressed as RDF triples, and the inference rules are expressed in N3 [60] format. The basic information regards the location, relations and capability of the persons. In order to simplify the discussion, distances between persons are given rather than generated in real-time by calling a service. Lastly, a few inference rules are used to infer relations between two persons.

```
#T0, fall detected
((action:confirm_fall action:check_assistance action:schedule_ambulance) (care:ambulance_service)).
((action:confirm_fall action:check_assistance action:schedule_assistance) (data:person_b)).
((action:confirm_fall action:check_assistance action:schedule_assistance) (data:person_c)).
((action:confirm_fall action:check_assistance) (care:Not_Needed)).
((action:confirm_fall) (care:Not_Needed)).

#T1, fall confirmed
(action:check_assistance action:schedule_ambulance) (care:ambulance_service)).
(action:check_assistance action:schedule_assistance)(data:person_b)).
(action:check_assistance action:schedule_assistance)(data:person_c)).
(action:check_assistance) (care:Not_Needed)).

#T2, (non-medical) assistance is required, with updated distance limit
((action:schedule_assistance) (data:person_c)).
```

Figure 7. Generated paths

Figure 7 shows at different time frames the generated paths that may lead to the satisfying target state, such that the fall event be resolved. Each row represents a possible path. In each row, the elements in the first list indicate the actions that path consists of. The elements in the second list specify the service providers who have been contracted to provide the assistance. The paths are generated with an N3 reasoner, EYE [56], by running a path generation engine[4] developed with the weighted state transition logic. EYE is a generic N3 reasoner that is applicable to a range of contexts. It can tackle any problem domain modelled in RDF (and N3 for rules) [57]. Both service descriptions stated in Figure 5, and the background data and knowledge stated in Figure 6 are used in the process of path generation, together with a query that states that the target is to get the fall event resolved.

---

[4] https://github.com/hongsun502/wstLogic/tree/master/engine

At T0, a fall event is detected, and there are five possible paths that may lead to the satisfying target state. Particularly, if the "confirm fall" action returns "fall is not confirmed," or if the "check assistance" action returns "no need for assistance," then there will be no need to provide assistance. All the possible paths at T0 start with the action to check if the fall is confirmed. While both person_b and person_c are considered as possible service providers in two separate paths, person_d is not considered in any of the possible paths. This is because it is not explicitly stated that person_d is capable of providing informal assistance, thus the precondition listed in the "schedule assistance" action fails. At T1, the state is updated: the event of fall is confirmed. There are four possible paths left: as can be observed, the path associated with a negative confirmation of the fall event no longer exists because the event is confirmed. Each of the four paths now start with checking whether assistance is needed, and does not check whether the fall event is confirmed anymore. At T2, the state is further updated with the fact that (non-medical) assistance is required. In conjunction, the precondition of the distance between the service provider and service receiver is also updated to be tighter. As a consequence, the only path left at T2 is to schedule assistance with person_c.

This fall event scenario shows that even a simple use case can be very complex in real-world application. In the demonstration we published on Github[5], we further extend this scenario to include diagnosis from EHRs and relevant clinical knowledge to deduce information and generate more appropriate paths. We also demonstrate how semantic gap prevents the utilization of available resources, as well as how we use semantic mapping to cope with such situations to build a solid descriptive intelligence. These scenarios demonstrate that our solution is capable of combining the descriptive, predictive, and prescriptive intelligence together, to generate adaptive and personalized carepaths, and construct a truly effective intelligent AAL system.

## 5. CONCLUSION

Ambient assisted living systems are developed with the aim to provide support to the elderly people and to enable their independent living at home. With the high prevalence of multiple chronic diseases, the elderly people are often in need of different levels of care management to prolong independent living at home, ranging from simple tasks such as taking medication, to more complicated ones such as adaptive and personalized carepath management. Most of the existing AAL systems are focusing on the function of assisted living, and lack support of care management. We deem that a truly effective AAL system should be an intelligent system that may serve as an extension of the care support provided in hospitals. With the development of the technologies that provide the relevant intelligence in the past decade, we consider it is now feasible to build such an intelligent AAL system.

---

[5] https://github.com/hongsun502/AALDemo/tree/main/fall_detection

This paper discussed the intelligence required in building an intelligent AAL system able to provide proper clinical support. Following the classifications originated in the data analytics domain, we propose that the intelligence required in AAL be classified as descriptive, predictive and prescriptive intelligence. We reviewed the literature related to the application of these three types of intelligence in the AAL domain, discussed the challenges and provided our views about the available solutions. Although we introduced these three types of intelligence separately, we consider that an intelligent AAL system should organically combine the power of these three types of intelligence together. We presented our views of building an AAL system with these three types of intelligence, based on semantic web technology as backbone to represent data, knowledge and services in a common standard and the use of weighted state transition logic to generate adaptive and personalized carepaths to provide clinical support to the elderly people.

## Acknowledgement

Hong Sun would like to acknowledge the GPS4IC project in supporting the development of the weighted state transition logic for adaptive carepath management. The authors would like to thank Jos De Roo for providing feedback on the demonstration of our scenario.

[3] Gingras, G., Adda, M., Bouzouane, A., Ibrahim, H., & Dallaire, C. (2020). IoT Ambient Assisted Living: Scalable Analytics Architecture and Flexible Process. Procedia Computer Science, 177, 396-404.

[4] Steg, H., & Strese, H. (2005). Ambient assisted living–european overview report.

[5] Huch, M., & Strese, H. (2005). Ambient Assisted Living-Preparing an Article 169 Measure. MST NEWS, 5, 8.

[6] AAL program, http://www.aal-europe.eu/, [Online; accessed February-2021].

[7] Tao, M., Ota, K., & Dong, M. (2017). Ontology-based data semantic management and application in IoT-and cloud-enabled smart homes. Future generation computer systems, 76, 528-539.

[8] Ni, Q., Pau de la Cruz, I., & Garcia Hernando, A. B. (2016). A foundational ontology-based model for human activity representation in smart homes. Journal of Ambient Intelligence and Smart Environments, 8(1), 47-61.

[9] Pardo, E., Espes, D., & Le-Parc, P. (2016). A framework for anomaly diagnosis in smart homes based on ontology. Procedia computer science, 83, 545-552.

[10] Perez, M. V., Mahaffey, K. W., Hedlin, H., Rumsfeld, J. S., Garcia, A., Ferris, T., ... & Turakhia, M. P. (2019). Large-scale assessment of a smartwatch to identify atrial fibrillation. New England Journal of Medicine, 381(20), 1909-1917.

[11] Hahnen, C., Freeman, C. G., Haldar, N., Hamati, J. N., Bard, D. M., Murali, V., ... & van Helmond, N. (2020). Accuracy of Vital Signs Measurements by a Smartwatch and a Portable Health Device: Validation Study. JMIR mHealth and uHealth, 8(2), e16811.

[12] Kang, M., Park, E., Cho, B. H., & Lee, K. S. (2018). Recent patient health monitoring platforms incorporating internet of things-enabled smart devices. International neurourology journal, 22(Suppl 2), S76.

[13] Basatneh, R., Najafi, B., & Armstrong, D. G. (2018). Health sensors, smart home devices, and the internet of medical things: an opportunity for dramatic improvement in care for the lower extremity complications of diabetes. Journal of diabetes science and technology, 12(3), 577-586.

[14] Haque, A., Milstein, A., & Fei-Fei, L. (2020). Illuminating the dark spaces of healthcare with ambient intelligence. Nature, 585(7824), 193-202.21